\documentclass[runningheads]{llncs}

\usepackage{algorithm}
\usepackage{algpseudocode}
\usepackage{amssymb}
\usepackage{amsmath}
\usepackage{graphicx}
\usepackage{subfig}
\usepackage{pgfplots}
\usepackage{soul}
\usepackage{cite}

\title{Parallel approach to sliding window sums} 

\usepackage{authblk}

\author{Roman Snytsar\inst{1} \and
Yatish Turakhia\inst{2}}
\authorrunning{R. Snytsar and Y. Turakhia}

\institute{Microsoft Corp., One Microsoft Way, Redmond WA 98052, USA
\email{Roman.Snytsar@microsoft.com} \and
Stanford University, Stanford CA 94305, USA}

\begin{document}

\maketitle

\begin{abstract}
{Sliding window sums are widely used in bioinformatics applications, including sequence assembly, k-mer generation, hashing and compression. New vector algorithms which utilize the advanced vector extension (AVX) instructions available on modern processors, or the parallel compute units on GPUs and FPGAs, would provide a significant performance boost for the bioinformatics applications.\\
We develop a generic vectorized sliding sum algorithm with speedup for window size $w$ and number of processors $P$ is $O(P/w)$ for a generic sliding sum. For a sum with commutative operator the speedup is improved to $O(P/log(w))$. When applied to the genomic application of minimizer based k-mer table generation using AVX instructions, we obtain a speedup of over 5$\times$.\\
}

\end{abstract}
\vspace{0.35cm}

\section{Introduction}
Bioinformatics algorithms for sequence assembly, indexing, search, and compression evolve at a breakneck rate. Still, many foundational ideas, like using short substrings known as k-mers, hashing schemas, and bespoke indexing mechanisms, stay relevant. A relatively new idea quickly gaining popularity is the use of minimizers \cite{Roberts04}. Minimizers exploit the sequence contiguity allowing to represent the sequence with a smaller number of k-mers, thus producing more compact indices. Minimizers have been successfully used for k-mer counting \cite{deorowicz15}, sequence alignment \cite{li18}, and indexing \cite{Wood14}. While some research has been directed towards improving the performance of minimizers \cite{marccais17}, little attention has been paid to the properties of the underlying algorithm of sliding window minimum.

In this paper we explore the properties of generic sliding window sums and uncover the potential for parallel speedup. We show a \emph{novel} sliding sum approach could be extended beyond minimizers to k-mer generation and hashing, and develop fast vector implementations not only for minimizer generation but also for k-mer hashing. In partiulcar, our approach provides a parallel speedup of $O(P/w)$ for a generic sliding sum with a window size $w$ and $P$ processors,  which could be further improved to a speedup of $O(P/log(w))$ for a sliding sum with a commutative operator. The rest of the paper is organized as follows. In section~\ref{sec:background}, we provide background for prefix sum and sliding sum algorithms, and their applications to the popular seed-filter-extend paradigm used in bioinformatics, particularly in the context of minimizers. In section~\ref{sec:method}, we present our generic vectorized sliding sum algorithm and show how it could be applied to our bioinformatics applications. We present our results in section~\ref{sec:results} and conclude in section~\ref{sec:conclusion}.


\section{Background}
\label{sec:background}

\subsection{Prefix Sum}

Parallel algorithms are often constructed from a set of universal building blocks. One of the hardest to identify, but extremely useful is the concept of a \emph{prefix sum}, and the accompanying \emph{scan} algorithm.
A prefix sum is a transformation that takes an operator $\oplus$, and a sequence of elements
\[ x_0, x_1, \ldots, x_k, \ldots \]
and returns the sequence
\begin{equation}\label{prefixSum}
y_i=\sum\limits_{j=0}^{i} x_j = x_0\oplus x_1\oplus\ldots\oplus x_i
\end{equation}

Despite the data carry dependency, the first $N$ elements of the prefix sum with an associative operator could be computed in $O(log(N))$ parallel steps using scan algorithm, as shown by \cite{Blelloch93}. 

\subsection{Sliding Window Sum}

Sliding window sum (sliding sum) takes a window size $w$ in addition to an operator $\oplus$, and a sequence of elements, and returns the sequence
\begin{equation}\label{slidingSum}
y_i=\sum\limits_{j=i}^{i+w-1} x_j = x_i\oplus x_{i+1}\oplus\ldots\oplus x_{i+w-1}
\end{equation}
where each sum is defined in terms of the operator $\oplus$ and contains exactly $w$ addends.
The asymptotic complexity of a naive sliding sum algorithm is $O(wN)$ where $N$ is the length of the source sequence.

It is worth mentioning that every sum defined by Equation~\ref{slidingSum} is a prefix sum with operator $\oplus$ and input sequence $x_i\ldots\oplus x_{i+w-1}$. Many useful operators are associative, so the prefix scan algorithm is applicable here, reducing complexity of every sum in Equation~\ref{slidingSum} to $O(log(w))$ and, trivially, the overall sliding sum complexity to $O(Nlog(w))$ parallel steps. We have observed this optimization being implemented by the vectorizing compilers. It is, however, possible to further improve performance by exploiting more subtle operator properties.

\subsection{The seed-filter-extend paradigm}

Most heuristics to local sequence alignment are based on the \emph{seed-filter-extend} paradigm, which was first popularized by the BLAST algorithm~\cite{ALTSCHUL90}. In aligning a reference sequence $R$ with a query sequence $Q$, the \textit{seeding} stage finds small local matches, called \emph{seed hits}, of size $k$ (also called k-mer, typically 10-19 base-pairs in size) between $R$ and $Q$. The \textit{filtering} stage itself may consist of several smaller sub-stages, which further reduces the search space by a combination of techniques, such as ungapped extension~\cite{ALTSCHUL90, harris2007improved} or chaining multiple seed hits in a diagonal band~\cite{li18, turakhia18}. The \textit{extension} stage typically performs the compute-intensive dynamic programming step, usually employing the Smith-Waterman equations~\cite{smith1981identification}.

\subsection{Seed tables and minimizers}

\begin{figure}
\centering
\includegraphics[width=0.9\textwidth]{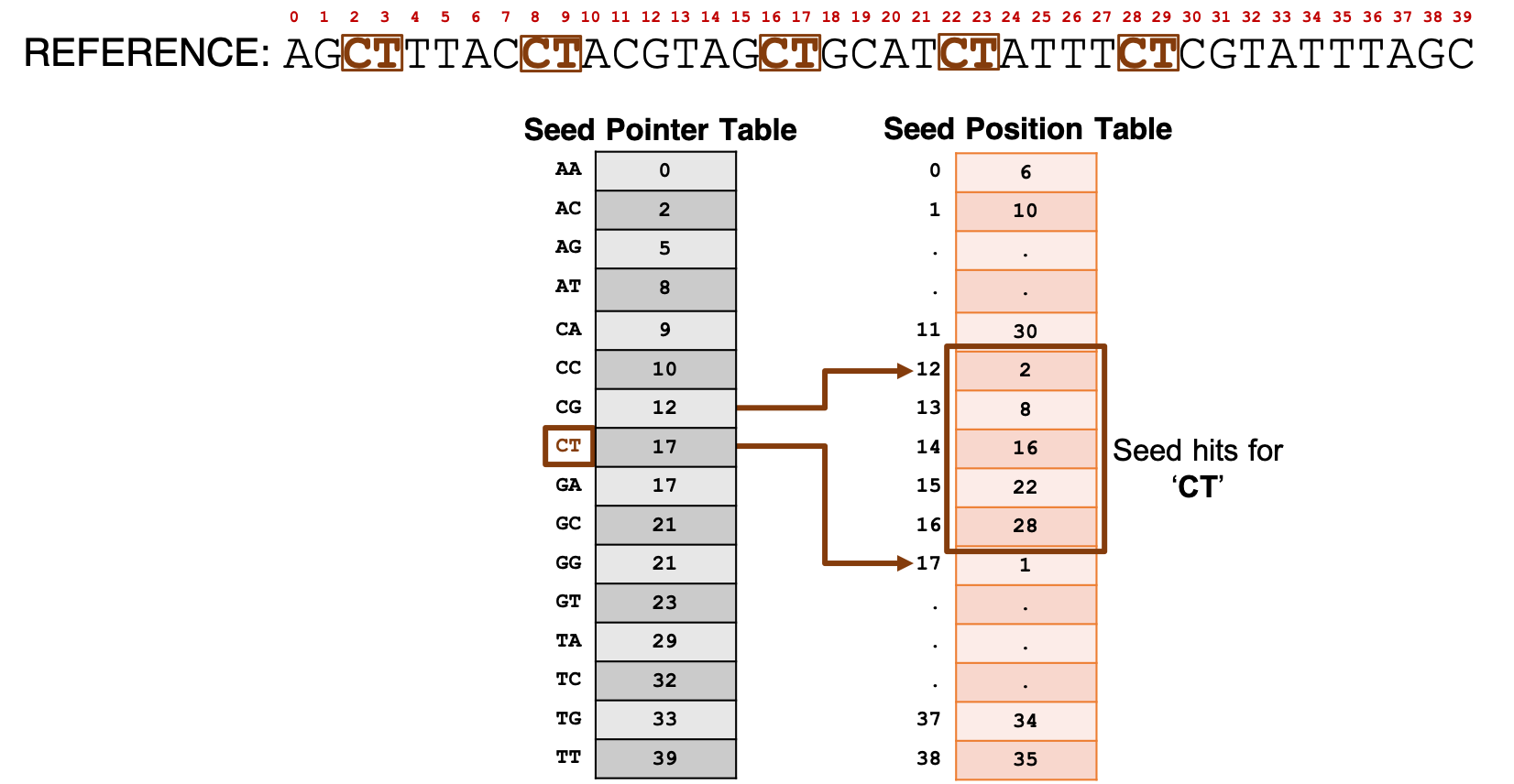}
\caption{An example reference sequence and seed table used in D-SOFT.
}
\label{fi:seed_table}
\end{figure}

Heuristics based on the seed-filter-extend paradigm often maintain a \emph{seed table} --- a data structure that enables fast lookup of seed hits in reference, $R$. Figure~\ref{fi:seed_table} shows an example reference sequence and seed table for seed size $k$=2. Seed table maintains two tables: (i) a seed pointer table and (ii) a seed position table. 
For each of the $4^k$ possible seeds (16 seeds in Figure~\ref{fi:seed_table}), lexicographically sorted, the seed pointer table points to the beginning of a list of hits in the seed position table. 
In Figure~\ref{fi:seed_table}, lookups to `CG' and `CT' in the seed pointer table give the start and end addresses in the seed position table for hits of `CT' in the reference. 

Starting with $R = {r_0, r_1, ... r_n}$, we can define k-mers of $R$ as a  sliding sum over window size k, string concatenation operator, and $R$.

\emph{Minimizer seeds} (or \emph{minimizers} for short), an idea originally proposed for compressing large seed tables in 2004 by Roberts et al.~\cite{Roberts04}, have seen a recent revival in bioinformatics with the advent long read alignment~\cite{li18} and metagenomics~\cite{wood2014kraken}. Minimizers can greatly reduce the storage requirements for the seed position table by storing only a subset of the seeds with only a small drop in sensitivity of the aligner. 

\begin{figure}
        \centering
        \subfloat[][]{
                \includegraphics[width=0.35\textwidth]{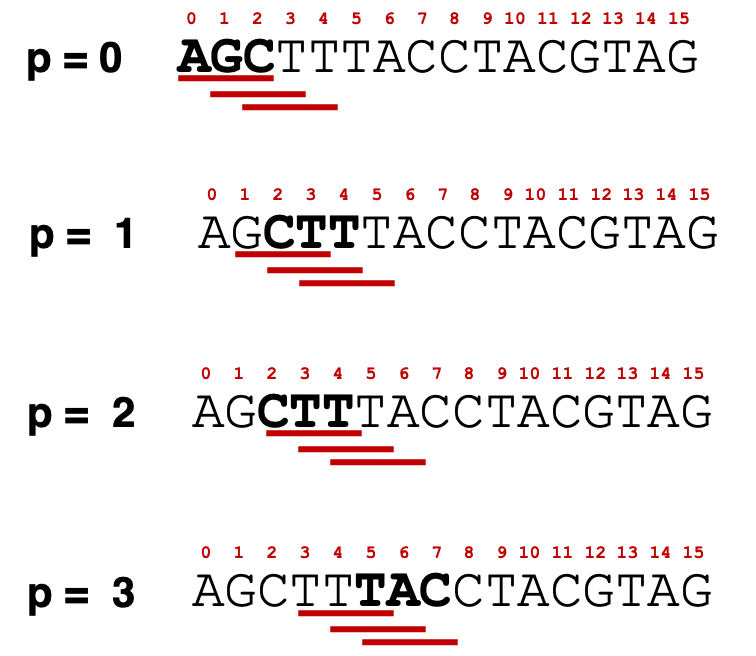}
                \label{fi:minimizer-1}
        }
        \subfloat[][]{
               \includegraphics[width=0.35\textwidth]{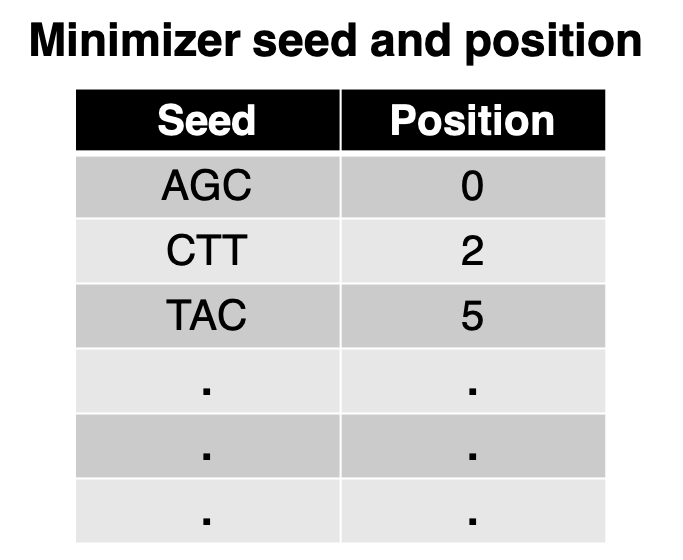}
                \label{fi:minimizer-2}
        }
        \caption{Illustration of minimizer seeds using ($k$=3, $w$=3). (a) An example reference sequence with a minimizer window sliding over 4 positions. The three seeds within the window are underlined in red and the minimizer seed within the window is highlighted in bold. (b) Minimizer seed-position pairs as constructed from (a).}
        \label{fi:minimizer}
\end{figure}

Figure~\ref{fi:minimizer} illustrates how minimizers can be used to build a seed position table with an example. In addition to the seed size $k$, minimizers require a parameter $w$, the \emph{minimizer window size}. In Figure~\ref{fi:minimizer}, $k=3$ and $w=3$. In each position $p$ of the reference $R$, a window $w$ consecutive seeds of size $k$ (k-mers) starting from position $p$ in $R$ are used to find the lexicograpically minimum seed $s$ and its position $p'$, which is recorded in the seed position table. Adjacent windows can share the same minimizer (i.e. the ($s$, $p'$) pair), which reduces the storage requirement for the seed position table. Figure~\ref{fi:minimizer-1} shows the minimizers for four consecutive positions 0-3 in $R$ and the corresponding entries in the seed position table in Figure~\ref{fi:minimizer-2}. Windows at positions $p=1$ and $p=2$ share the same minimizer (`CTT', 2), which is stored only once in Figure~\ref{fi:minimizer-2}. Moreover, as seen in Figure~\ref{fi:minimizer-2}, seeds at position 0, 2 and 5 are stored in the seed position table but those at positions 1, 3 and 4 are dropped. Roberts et al.~\cite{Roberts04} have shown that with a minimizer window of size $w$, a new minimizer occurs every $w/2$ bases on average. 

Minimizers are a key innovation in Minimap~\cite{li2016minimap} and its successor Minimap2~\cite{li18}, both of which achieve an order of magnitude speedup over prior techniques, most speedup resulting from fewer seed hits per read due to minimizers. We have found that turning off minimizers (using $w=1$ instead of the default $w=10$) slows down the seeding and filtering stage of Minimap2 by nearly 7$\times$ with only 0.5\% higher sensitivity for sequencing reads from Pacific Biosciences. As evident from figure~\ref{fi:minimizer}, minimizer seed table construction is a form of sliding window sum with operator $min$ over a window of k-mers, requiring $O(wN)$ for construction. It is possible to achieve $O(N)$ complexity at the cost of using elaborate queue-based data structures \cite{tangwongsan15}. Constructing seed tables can take several hours for the \emph{de novo} assembly of a human genome~\cite{turakhia18}. In this paper, we take a closer look at the connection between sliding sums and prefix sums, and attempt to supersede the linear complexity achieved by previous approaches.

\section{Methods}
\label{sec:method}

\subsection{Vector Algorithms}

Our first algorithm is a vector-friendly way of calculating sliding sum assuming the input sequence elements become available one by one and are processed using the vector instructions of width $P>w$:
\begin{algorithm}
	\caption{Scalar Input}\label{al:scalar}
	\begin{algorithmic}[0]
		\Procedure{ScalarInput}{$x_0\dots x_{n-1}$}
		\State $Y\gets \Big(\underbrace{\sum\limits_{j=0}^{w-2} x_j, \sum\limits_{j=1}^{w-2} x_j, \dots , x_{w-3}\oplus x_{w-2}, x_{w-2}}_{w-1}, 0, \dots , 0\Big)$
		\For{$i=w-1$ to N}
		\State $X\gets \Big(\underbrace{x_i, x_i, \dots , x_i}_w, 0, \dots , 0\Big)$
		\State $Y\gets Y \oplus X$
		\State $y_{i-w+1}\gets Y[0]$
		\State $Y\gets Y \lll 1$
		\EndFor
		\EndProcedure
	\end{algorithmic}
\end{algorithm}
\begin{figure}
\centering
\includegraphics[width=1.0\textwidth]{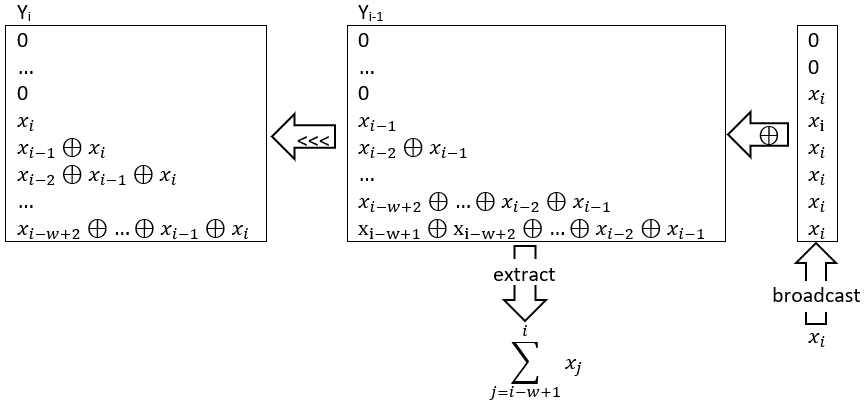}
\caption{Data flow of the scalar input sliding sum algorithm.
}
\label{fi:algo1}
\end{figure}

Vector Y is initialized to the suffix sums with the number of elements decreasing from $w-1$ to $0$. Then in a loop every incoming element $x_k$ is broadcast to the first $w$ elements of vector X. After vector addition the zeroth element of Y contains the next sliding sum. Next, the vector Y is shifted left by one element, as denoted by operator $\lll$, and the state is ready for the next iteration. The data flow of the scalar algorithm is depicted on the Figure \ref{fi:algo1}

Asymptotic complexity of the scalar input algorithm is $O(N)$ with no additional requirements on the operator $\oplus$.

This result could be improved if we assume that the input sequence arrives packed in vectors of width $P>w$.
\begin{algorithm}
\caption{Vector Input}\label{al:vector}
\begin{algorithmic}[0]
\Procedure{VectorInput}{$x_0\dots x_{n-1}$}
\State $Y\gets \Big(\underbrace{\sum\limits_{j=0}^{w-2} x_j, \sum\limits_{j=1}^{w-2} x_j, \dots , x_{w-3}\oplus x_{w-2}, x_{w-2}}_{w-1}, 0, \dots , 0\Big)$
\For{$i=w-1$ to N step P}
\State $X\gets \Big(x_k, x_{k+1}, \dots , x_{k+p-1}\Big)$
\State $X1\gets \Big(\underbrace{X_0, X_0\oplus X_1, \dots , \sum\limits_{j=0}^{w-2} X_j}_{w-1}, \sum\limits_{j=0}^{w-1} X_j, \dots , \sum\limits_{j=p-w}^{p-1} X_j\Big)$
\State $Y1\gets \Big(0, \dots , 0, \underbrace{\sum\limits_{j=p-w}^{p-1} X_j, \sum\limits_{j=p-w}^{p-2} X_j, \dots , X_{p-w}}_{w-1} \Big)$
\State $Y\gets Y \oplus X1$
\State $ y_{k-w+1}\dots y_{k-w+p}\gets Y[0]\dots Y[p-1]$
\State $Y\gets Y1 \lll (P-w)$
\EndFor
\EndProcedure
\end{algorithmic}
\end{algorithm}

\begin{figure}
\centering
\includegraphics[width=1.0\textwidth]{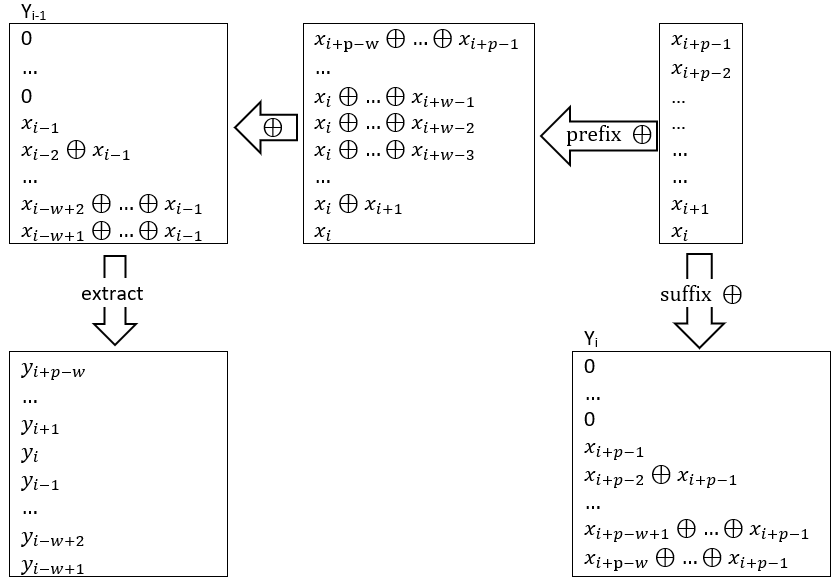}
\caption{Data flow of the vector input sliding sum algorithm.
}
\label{fi:algo2}
\end{figure}

At every iteration $P$ input elements are placed into vector $X$. X1 is filled with the prefix sums of up to $w$ addends, and Y1 is filled with the suffix sums constructed from the elements of $X$, as shown on the Figure \ref{fi:algo2}. Then the vector sum of $Y$ and $X1$ yields the next P output elements. Finally, the suffix sums from $Y1$ are shifted into proper positions in vector $Y$, and it is ready for the next iteration. 

The asymptotic complexity thus is  $O(N\cdot w/P)$ with the parallel speedup $O(P/w)$ for any operator $\oplus$. If $\oplus$ is associative, the prefix/suffix sums could be computed in parallel using the algorithm in \cite{Blelloch93}, and the complexity is reduced to $O(N\cdot log(w)/P)$ with the speedup improving to $O(P/log(w))$.

For example, since $min$ is an associative operator, the sliding window minimum can be computed using the faster version of the vector input algorithm.

\section{Results}
\label{sec:results}

\begin{figure}[!tpb]
\centering
	\begin{tikzpicture}
	\begin{axis}[
	xlabel=Window Size,
	ylabel=Time\, s,
	legend style={at={(0.8, 0.4)},anchor=north}]
	\addplot[mark=square*] coordinates {
		(4,71.00)
		(5,74.93)
		(6,79.88)
		(7,79.10)
		(8,79.46)
		(9,78.83)
		(10,79.03)
		(11,79.13)
		(12,79.55)
		(13,80.12)
		(14,80.79)
		(15,79.69)
		(16,79.74)
		(17,79.94)
	};
	\addplot[mark=x] coordinates {
		(4,61.84)
		(5,67.09)
		(6,66.56)
		(7,65.80)
		(8,53.87)
		(9,55.90)
		(10,56.26)
		(11,61.64)
		(12,61.54)
		(13,65.29)
		(14,68.72)
		(15,74.16)
		(16,53.62)
		(17,55.43)
	};
	\addplot[mark=*] coordinates {
		(4,16.88)
		(5,16.94)
		(6,14.07)
		(7,14.11)
		(8,13.97)
		(9,14.05)
		(10,11.28)
		(11,11.21)
		(12,11.16)
		(13,10.95)
		(14,11.31)
		(15,11.26)
		(16,11.23)
		(17,11.26)
	};
	\legend{Dequeue, Array, Vector}
	\end{axis}
	\end{tikzpicture}
	\caption{Performance of the sliding minimum algorithms.}\label{minperf}
\end{figure}
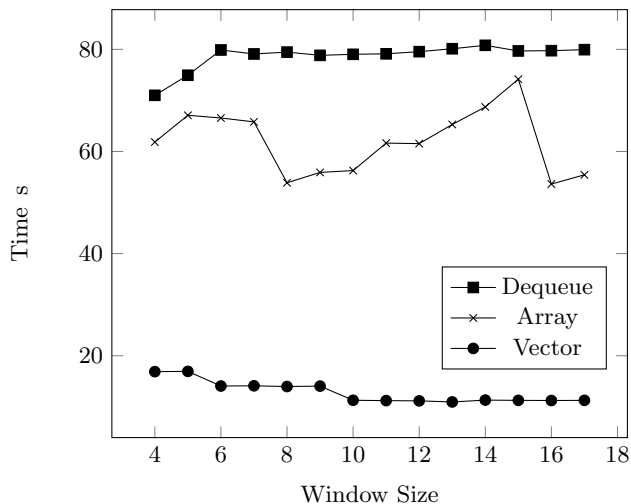

We tested the performance of various sliding minimum algorithms using the hashed 15-mers of the reference human genome assembly (GRCh38) from the Genome Reference Consortium. The test imitates a minimizer based seed table construction by a long-read aligner, such as Minimap2 \cite{li18}, GraphMap \cite{sovic16} or Darwin \cite{turakhia18}. Figure {\ref{minperf}} compares the performance of the na\"ive array-based algorithm, linear dequeue-based algorithm, and our proposed vector algorithm.

Deque-based algorithm performance is indeed independent of the window size. It comes, however, at the cost of a significant overhead of managing the deque data structure and unpredictable branching.

Array-based algorithm, despite the worst asymptotic complexity, is simple to implement, and benefits from the automatic compiler vectorization. It is clear how the times drop when the window size is aligned with the SIMD vector width ($P$ = 4, 8, and 16). For small window sizes the array algorithm is competitive with the deque approach.

Our vector sliding sum algorithm beats both previous implementations by a factor of $5\times$. With the SSE/AVX instruction set, any window size requires the same number of instructions as the closest (larger) power of 2. So the performance of our vector implementation does not change linearly with $w$ but drops when we switch to the different SIMD vector width $P$ at $w = 5, 9, 17$. Also, prefix sum computation across wider vectors incurs additional latencies for cross-lane data exchanges, resulting in the speedup less than theoretical $2\times$. 

\section{Conclusion}
\label{sec:conclusion}
We introduced a family of algorithms for parallel evaluation of sliding window sums. The parallel speedup for window size $w$ and number of processors $P$ is $O(P/w)$ for a generic sliding sum. For a sum with a commutative operator the speedup is improved to $O(P/log(w))$. For a family of sliding sums that allow recurrent interpretations, the speedup is independent of $w$: $O(P/log(P))$. This gives the developer a choice of fast branchless algorithms suitable for implementation on any modern parallel architecture including modern CPUs with instruction-level parallelism, pipelined GPUs, or FPGA reconfigurable hardware.

While we concentrate on the faster sliding window sum algorithms for bioinformatics, our findings are relevant for accelerating all the numerous sliding window applications from compression and cryptography to high frequency data mining \cite{ikonomovska07}.

\bibliographystyle{splncs04}


\bibliography{ParallelSlidingSum}
\end{document}